\documentclass[amsmath,amssymb,aps,prc,reprint,superscriptaddress]{revtex4-2}
\usepackage{booktabs}
\usepackage{graphicx}
\usepackage{float}
\usepackage{xcolor}
\usepackage{dcolumn}

\usepackage{bm}
\usepackage{hyperref}
\usepackage[T1]{fontenc}
\usepackage[utf8]{inputenc}
\usepackage[margin=0.8in]{geometry}

\usepackage{siunitx}
\usepackage{array}
\usepackage{longtable} 
\usepackage{tabularx}  
\usepackage{geometry}
\newcommand{{\GEANT}}{\textsc{Geant4}}
\newcommand{{\MENATER}}{\textsc{Menate\_r}}
\newcommand{{\MCNP}}{\textsc{Mcnp}}
\begin{document}

\title{Exclusive cross section measurements of  $^{12}$C(n,p$_\text{n}$) and $^{12}$C(n,d$_\text{n}$) reactions}

\author{A. Wantz}
  \email{wantz@frib.msu.edu}
\affiliation{Facility for Rare Isotope Beams, Michigan State University, East Lansing, Michigan, 48824}
\affiliation{Physics and Astronomy Department, Michigan State University, East Lansing, Michigan, 48824}

\author{A. N. Kuchera}
\affiliation{Department of Physics, Davidson College, Davidson, North Carolina, 28305}

\author{W. F. Rogers}%
\affiliation{Department of Physics, Indiana Wesleyan University, Marion, Indiana, 46953}

\author{T. Baumann}
\affiliation{Facility for Rare Isotope Beams, Michigan State University, East Lansing, Michigan, 48824}

\author{C. Boothby}
\affiliation{Department of Physics, Davidson College, Davidson, North Carolina, 28305}

\author{M. Devlin}%
\affiliation{Los Alamos National Laboratory, Los Alamos, New Mexico, 87545}

\author{N. Frank}
\affiliation{Augustana College, Rock Island, Illinois, 61201}

\author{P. Gueye}
\affiliation{Facility for Rare Isotope Beams, Michigan State University, East Lansing, Michigan, 48824}
\affiliation{Physics and Astronomy Department, Michigan State University, East Lansing, Michigan, 48824}

\author{K. J. Kelly}
\affiliation{Los Alamos National Laboratory, Los Alamos, New Mexico, 87545}

\author{S. A. Kuvin}%
\affiliation{Los Alamos National Laboratory, Los Alamos, New Mexico, 87545}

\author{A. LaRochelle}
\affiliation{Department of Physics, Davidson College, Davidson, North Carolina, 28305}

\author{T. Redpath}
\affiliation{Virgina State University, Petersburg, Virginia, 23806}

\author{T. Tran Nguyen}
\affiliation{Department of Physics, Davidson College, Davidson, North Carolina, 28305}

\author{J. R. Winkelbauer}
\affiliation{Los Alamos National Laboratory, Los Alamos, New Mexico, 87545}

\date{\today}

\begin{abstract}
A single-crystal chemical vapor deposited (sCVD) diamond detector was used at LANSCE to measure neutron-induced reactions on carbon. Neutrons with a broad energy spectrum were produced via spallation. Relative cross sections are reported from above threshold up to 26~MeV incident neutron energy for $^{12}$C(n,p$_2$)$^{12}$B, up to 23~MeV for $^{12}$C(n,p$_{3+4}$)$^{12}$B, and up to 35~MeV for $^{12}$C(n,d$_1$)$^{11}$B and $^{12}$C(n,d$_2$)$^{11}$B, all of which represent the highest-energy measurements for these reactions to date, and the first measurement for (n,d$_2$).  Cross sections for $^{12}$C(n,p)$^{12}$B$_\text{total}$ and $^{12}$C(n,d)$^{11}$B$_\text{total}$ are also reported from sums of the partial cross sections. These measurements are compared with recent data from EXFOR as well as TALYS calculations. These data are useful for future nuclear data evaluations and can be used as cross section inputs for data-based simulation packages.

\end{abstract}

\maketitle

\section{\label{sec:level1}Introduction }


Neutron-induced reactions on $^{12}$C are of longstanding interest in both fundamental and applied nuclear physics, requiring accurate partial cross section data over a broad energy range. However, nuclear data evaluations of reactions on $^{12}$C are hindered by a scarcity of partial cross section data, 
particularly for charged-particle-emitting reactions above 20 MeV. Additionally, challenges arise in microscopic calculations from the number of open reaction channels, many of which lead to particle-unbound states, complicating theoretical approaches.

Previous studies \cite{Kellogg1953,Kelly1950,Klug2003} have shown that a substantial portion of the total n-$^{12}$C cross section is due to inelastic reactions with $^{12}$C, which generally have not been measured exclusively, especially above 20 MeV. Although numerous studies have demonstrated the ability to measure n--$^{12}$C cross sections with a diamond detector \cite{wantz2025, Pillon2011, Pillon2017, MAJERLE2020163014, Kuvin2021, LIU2023137985}, only the works of Pillon \textit{et al}. \cite{Pillon2011,Pillon2017} have published cross sections for populating higher-lying states of $^{11,12}$B via $^{12}$C(n,p$_2$)$^{12}$B, $^{12}$C(n,p$_3$)$^{12}$B, $^{12}$C(n,p$_4$)$^{12}$B, and $^{12}$C(n,d$_1$)$^{11}$B. 

The present work extends the cross section measurements of these reactions to higher incident neutron energies,  leveraging data from Ref. \cite{wantz2025} to extract cross sections for four additional reaction channels: $^{12}$C(n,p$_2$)$^{12}$B, $^{12}$C(n,p$_{3+4}$)$^{12}$B, $^{12}$C(n,d$_1$)$^{11}$B, and $^{12}$C(n,d$_2$)$^{11}$B. Notably, this is the  first cross section measurement for $^{12}$C(n,d$_2$)$^{11}$B.
Combined with the data of Ref. \cite{wantz2025}, the sum of these partial cross sections reproduces the $^{12}$C(n,p)$^{12}$B and $^{12}$C(n,d)$^{11}$B total cross sections up to 17.3 MeV and 21.6 MeV, respectively. These measurements provide new data for nuclear data evaluations. In particular, the ENDF/B- evaluation for the $^{12}$C(n,p)$^{12}$B cross section can be updated with more recent measurements, at higher energies. These cross section measurements will be especially useful if evaluations are performed for exclusive channels.

In this work, TALYS \cite{TALYS} is used to provide a theoretical comparison with the data. 
Although statistical calculations like TALYS are not expected to be fully reliable in this mass and energy range, it is one of the only options for some level of a predictive calculation. For each reaction channel, TALYS calculates contributions from direct and compound reaction mechanisms, and the resulting total cross section. Comparisons with measured cross section data help identify deficiencies in the modeling of the direct and/or compound components, underscoring the need for new data to better constrain TALYS and other theoretical models. 
Thus, while TALYS should by no means be considered the definitive physics model, it is useful as a representative theoretical comparison. Although the default TALYS calculations do a reasonably good job of getting the scale for most reactions correct, the current limitations of the model are apparent throughout this manuscript.

This work was also motivated by the need to improve simulation of nuclear physics experiments using organic scintillators. These materials are commonly used to detect fast neutrons and are composed of carbon and hydrogen atoms. Many widely used software packages for simulating neutron interactions with these detectors (including \GEANT $ $ \cite{AGOSTINELLI2003250,ALLISON2016186,GEANT4_2nd_reference}, \MENATER $ $ \cite{Roeder}, \MCNP $ $ \cite{TechReport_MCNP}, and neuSIM4 \cite{PARK2024169475}) use an empirical approach, necessitating accurate cross section data for simulation inputs. With rare isotope facilities providing access to neutron rich and neutron unbound nuclei, accurate simulation tools are critical for planning and analyzing experiments. 



An overview of the experiment is included in Sec. \ref{exp}. Key aspects of the cross section normalization are presented in Sec. \ref{flux determination}, including background subtraction (Sec. \ref{bkg}), efficiency correction (Sec. \ref{eff}), and uncertainty quantification (Sec. \ref{uncertainty}). The cross section results are presented in Secs. \ref{np2}-\ref{nd2} for (n,p$_2$), (n,p$_{3+4}$), (n,d$_1$), and (n,d$_2$), respectively. The $^{12}$C(n,p)$^{12}$B and 
$^{12}$C(n,d)$^{11}$B total cross sections are compared to TALYS and previous experimental data, including ENDF/B-VIII.0, in Secs. \ref{np_total} and \ref{nd_total}. 
 
\section{Experiment} \label{exp}
Protons were accelerated to 800 MeV and impinged on a tungsten spallation target (Target 4) at the Los Alamos Neutron Science Center (LANSCE) \cite{Lisowski1991}. This produced fast neutrons with a broad energy distribution from 0 to above 600 MeV, with 1.8 $\mu$s separating each group of neutrons due to the pulse structure of the proton beam. The neutron beam was collimated to a 3 mm diameter beam spot entering the 90 m station on the 15L flight path, in which a Cividec B8 diamond detector \cite{Cividec} with 4$\times$4 mm$^2$ active area was installed. This detector used the Cx spectroscopic shaping amplifier \cite{Cividec}. Due to its high gain, the signals were attenuated by a factor of 10 before being routed into a CAEN VX1730B digitizer with pulse shape discrimination firmware. Only the waveform integrals, rather than the waveforms themselves, were collected in the data acquisition. A second diamond detector was placed downstream with a C6 fast charge amplifier \cite{Cividec}, but was not used in this analysis, due to its lower energy resolution. These semiconductor detectors detect neutron-induced charged reaction products, such as protons, deuterons, and recoiling carbon nuclei, which ionize carbon atoms in the detector volume, creating electron-hole pairs that drift to the electrodes under the applied electric field. This produces signals with good energy and timing resolution. For further details about the experimental setup, including a diagram of the flightpath, refer to Ref. \cite{wantz2025}.  

\begin{figure*}[t!]
\includegraphics[width=0.74\textwidth]{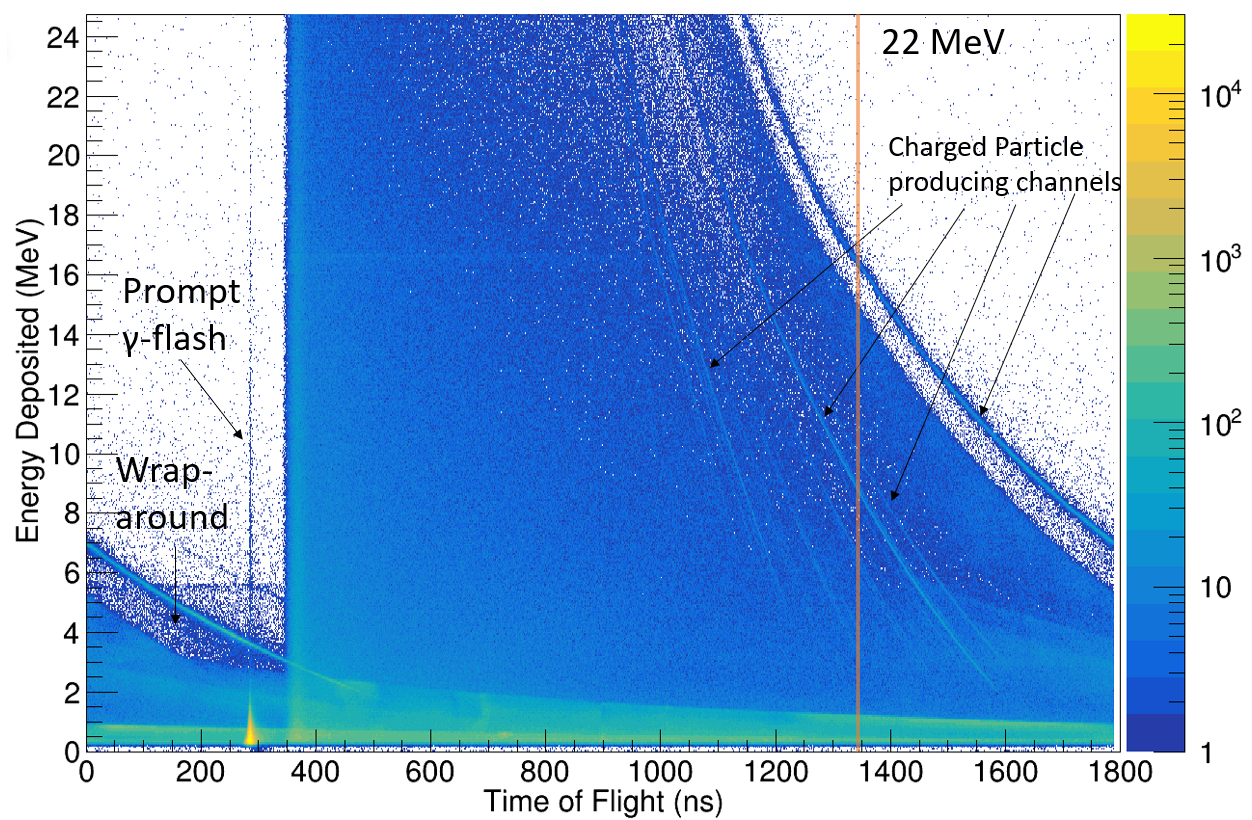}
\caption{\label{fig:energyvstof} Energy detected in the diamond detector as a function of time of flight from the spallation target to the diamond detector. Notable features include the $\gamma$ flash, the wraparound neutron background, and some well-defined bands corresponding to reaction channels producing only charged particles. The vertical orange line corresponds to a time of flight region centered on 22 MeV (see Fig. \ref{fig:qval}).} 
\end{figure*}

The timing spectra for the detectors are measured relative to a proton pick-off detector upstream of the spallation neutron source.  The neutron time of flight was then calibrated relative to the arrival of prompt gamma-rays, which provide a zero-time reference, allowing for the conversion of neutron time of flight to incident neutron kinetic energy (E$_n$). The energy deposited in the diamond as a function of the neutron time of flight is shown in Fig. \ref{fig:energyvstof}.

\begin{figure*}
\includegraphics[width=2\columnwidth]{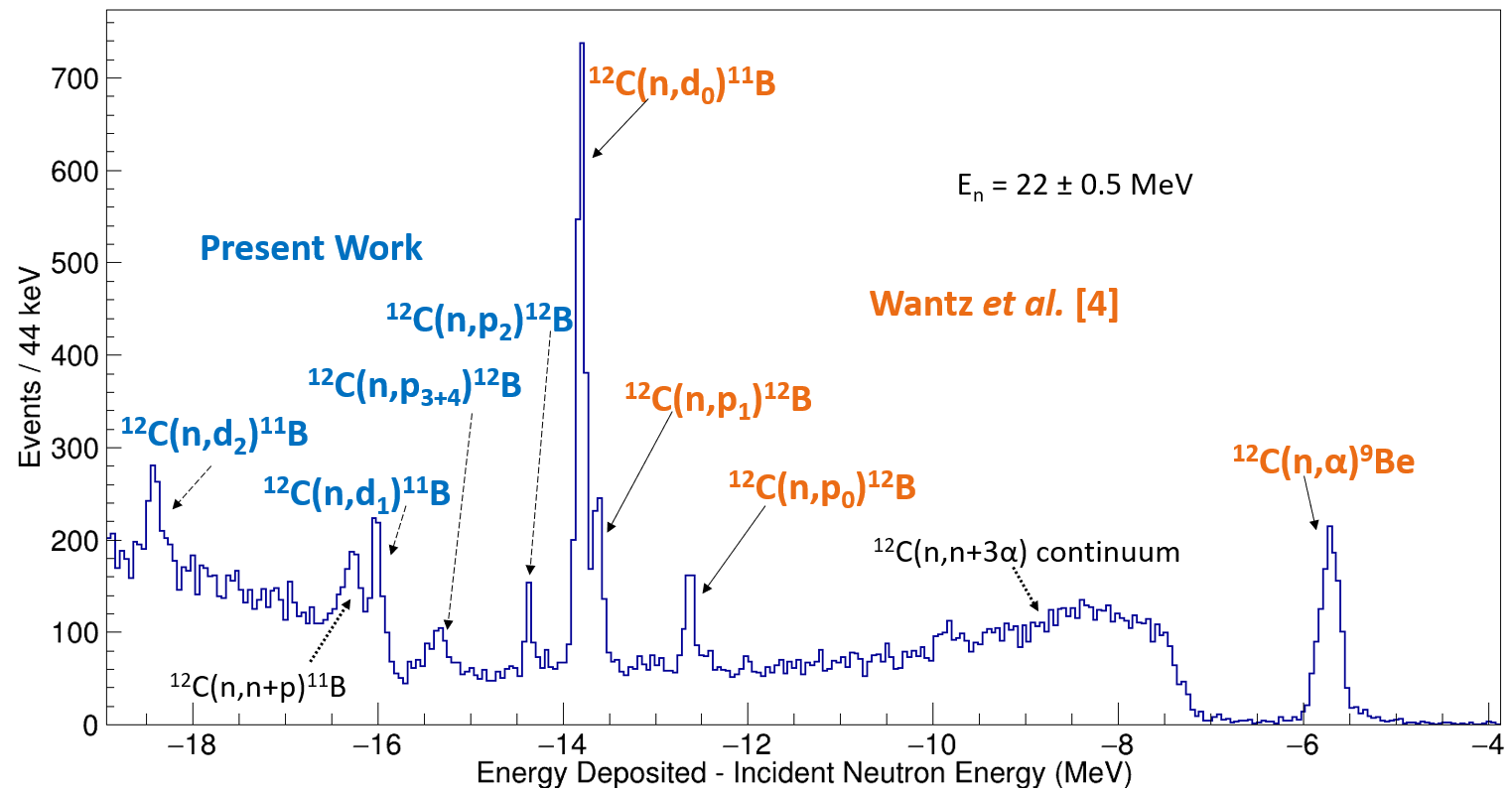} 
\caption{\label{fig:qval} Reconstructed reaction Q-value spectrum for 22 MeV neutrons, allowing for identification of peaks corresponding to different reaction channels. The data in this spectrum correspond to the shaded region around 1350 ns in  Fig. \ref{fig:energyvstof}. The reaction channels featured in this work are labeled in blue text, while those studied in Ref. \cite{wantz2025} are labeled in orange. The reader is referred to the online version for color figures.}
\end{figure*}

The data can be visualized by plotting the reconstructed reaction Q value, which is calculated by subtracting the incident neutron energy from the energy deposited in the detector. An example of a reconstructed Q value spectrum is plotted in Fig. \ref{fig:qval}. The four reaction channels for which cross sections were obtained in this work are labeled, along with four reaction channels previously measured \cite{wantz2025} from the same dataset. This labeling follows the convention used in recent literature \cite{Kuvin2021, Pillon2011, Pillon2017, MAJERLE2020163014, wantz2025}, in which the level of the daughter nucleus is denoted as a subscript on the light ion, so that the shorthand notation uniquely identifies each reaction. For example, $^{12}$C(n,p$_2$)$^{12}$B, referring to production of a proton and $^{12}$B in the second excited state, can be abbreviated as (n,p$_2$).  

In addition to the four peaks for which cross sections were determined in this study, another peak was identified, corresponding to the $^{12}$C(n,n+p)$^{11}$B reaction. Although in Fig. \ref{fig:qval} this appears as a fairly sharp peak, at other energies its appearance more closely resembles that of the $^{12}$C(n,n+3$\alpha$) continuum. This makes it difficult to reliably extract the yield; therefore, cross sections for (n,n+p) are not presented in this work. 

\section{Normalization} \label{flux determination}

\subsection{Neutron flux determination} 
The relative neutron flux distribution was measured with a fission chamber (see Fig. 4 in Ref. \cite{wantz2025}). The cross sections were normalized to previous (n,$\alpha$) cross section data at 14.1 MeV, by the same method described in Refs. \cite{wantz2025, Kuvin2021}.


\subsection{Background}\label{bkg}
The external background is negligible, due to the small size of the detector. The internal background is non-trivial, due to neutron interactions in the diamond which do not deposit their full energy. Various processes can contribute to these events, including reactions with one or more neutrons in the exit channel, such as $^{12}$C(n,n)$^{12}$C  (elastic scattering), $^{12}$C(n,n+3$\alpha$), $^{12}$C(n,2n)$^{11}$C, and $^{12}$C(n,n+p)$^{11}$B. Some of the energy is not measured because it is carried away by these outgoing neutrons, leading to an inaccurate Q value reconstruction. 
Additionally, light charged particles produced from neutron interactions can have sufficient kinetic energy that they escape the detector; such events lead to an incomplete energy-deposition tail for each of the peaks, which is more significant at higher incident neutron energies. These events, combined with $\gamma$ and $\beta$ decay of unstable nuclei, contribute to an approximately linear background which affected every channel studied in this work.  To extract the counts, each peak in the Q-value spectrum was fit with a Gaussian plus a linear background, which generally reproduces the data well. This operates under the assumption that the background does not exhibit sharp changes in magnitude within the width of a peak, which has not been observed in this work. The counts from the peak (background subtracted) were used in the cross section calculation. To conservatively account for any other uncertainties arising from the fitting process, an additional fit uncertainty was added, which is taken from the uncertainty in the width of the peak for each fit. This uncertainty is quantified in the supplemental material \cite{supp}.

\subsection{Charged-particle efficiency correction}\label{eff}
\begin{figure}
\includegraphics[width=\columnwidth]{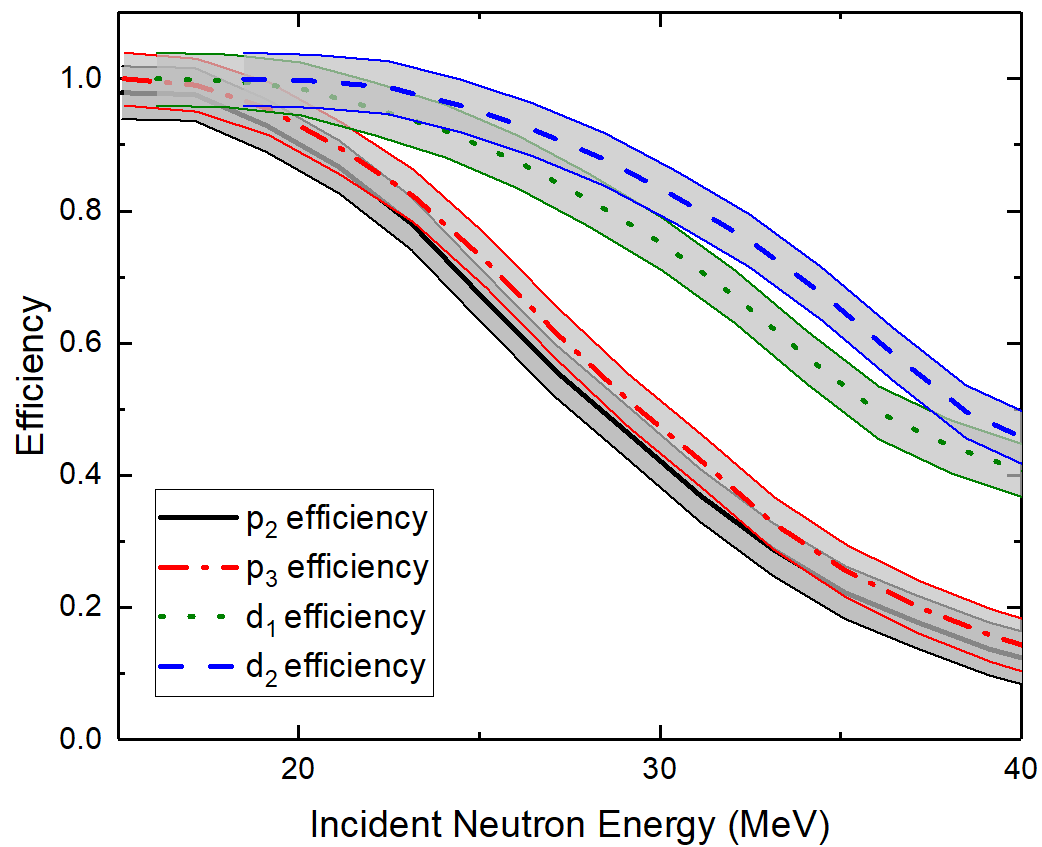}
\caption{\label{fig:Efficiency} Simulated detection efficiency as a function of incident neutron energy for each of the reactions studied. The curves correspond to the detection efficiency for the products of each of the reactions,  $^{12}$C(n,p$_2$)$^{12}$B, $^{12}$C(n,p$_3$)$^{12}$B, $^{12}$C(n,d$_1$)$^{11}$B, and $^{12}$C(n,d$_2$)$^{11}$B, respectively. The efficiencies were used to correct the cross sections. Uncertainty bands of $\pm$4\% are included around the simulated values.}
\end{figure}

The diamond detectors used in this experiment are thin, with a thickness of 0.5 mm. As a result, reaction products often escape the detector volume without depositing their full energy. To estimate the probability for reaction products to escape the detector, simulations were run with \MENATER $ $ in \GEANT $ $. These simulations were used to estimate the likelihood for events to show up in the full energy peak, which corresponds to an accurate identification experimentally in the Q-Value spectrum, as opposed to events where the particles escape, which lead to an incomplete energy-deposition tail in the Q-Value spectrum. These simulations use an isotropic angular distribution in the CM frame. A previous study \cite{Kuvin2021} determined the uncertainty arising from the use of these angular distributions rather than measured angular distributions (which for many of the reaction channels in question have not been measured) to be no larger than 4\%. The results of these simulations are shown in Fig. \ref{fig:Efficiency}. The yields for each reaction channel were corrected by the corresponding detection efficiencies.

\subsection{Uncertainty}\label{uncertainty}
The total uncertainties for the cross section measurements include both statistical and systematic contributions, but are generally dominated by statistical uncertainties, typically between 10-30\%. Several systematic uncertainties also affect the total uncertainty. An uncertainty of 4.9\% arises from the previous $^{12}$C(n,$\alpha$)$^9$Be data \cite{Pillon2011, Schmidt1992,Haight1984, Kondo2008, SANAMI2000403} that was used for cross section normalization. Differences between the angular distributions used in \MENATER $ $ and measured angular distributions, or the lack of measured angular distributions for some channels, result in a 4\% uncertainty in the charged particle detection efficiencies. Uncertainties in the neutron flux distribution used for normalization start at 5\% and gradually increase with increasing incident neutron energy. An additional uncertainty, typically below 10\%, was adopted to account for fitting the peaks and background to extract the counts, although the uncertainties increase for the highest-energy cross sections. Due to the long flight path length and good timing resolution of the detector, the maximum E$_n$ reconstruction uncertainty was 0.49\%, and was lower for all other cross sections in this work. The horizontal error bars in the cross section plots were obtained from this uncertainty. The total uncertainty for each cross section was obtained from the quadrature sum of the listed uncertainties. These uncertainties are fully enumerated in the supplemental material \cite{supp}. The uncertainties for these channels are generally larger than those of Ref. \cite{wantz2025}, driven by the smaller cross sections and hence lower statistics.

Some uncertainties are correlated for different data points within this work (energy-energy correlations). In  order of most to least impact on the total uncertainty: The flux uncertainty is partially correlated. The fit uncertainty is uncorrelated. The normalization uncertainty is fully correlated. The efficiency correction uncertainty is fully correlated. The neutron energy reconstruction uncertainty is fully correlated. In this work, the sources of uncertainty with lesser contributions to the total uncertainty are fully correlated, while the sources of uncertainty with greater contribution to the total uncertainty are uncorrelated.  

\section{Results}

\subsection{\boldmath  $^{12}$C(n,p$_2$)$^{12}$B} \label{np2}
\begin{figure}
\includegraphics[width=\columnwidth]{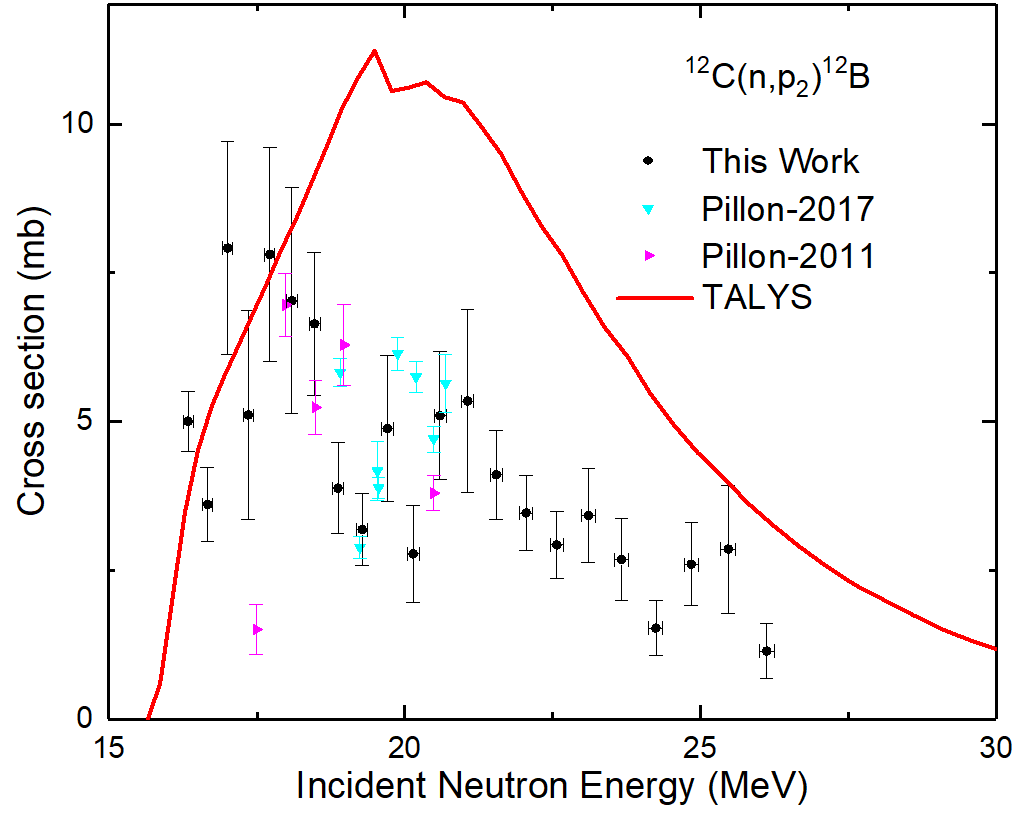}
\caption{\label{fig:np2_xs} Cross sections for $^{12}$C(n,p$_2$)$^{12}$B, normalized to previous $^{12}$C(n,$\alpha$)$^{9}$Be data. This work and previous data are compared with a TALYS calculation using default parameters.}
\end{figure}

Cross sections were measured for $^{12}$C(n,p$_2$)$^{12}$B, i.e. production of a proton and $^{12}$B in the second excited state (J$^\pi =$ 2$^-$, E$_x$ = 1.673 MeV), and are shown in Fig. \ref{fig:np2_xs} and Table \ref{tab:table_p2}. These measurements extend from E$_n =$ 16.34 to 26.11 MeV. The only previous cross section measurements for this channel were conducted by Pillon \textit{et al}. \cite{Pillon2011, Pillon2017}. Good agreement is found between the present results and those of and Ref. \cite{Pillon2011, Pillon2017}, although our uncertainties are much larger, primarily due to statistics. Some possible resonances are observed around  E$_n$ = 18 and E$_n$ = 21 MeV, though the fluctuations in the cross section are on a similar scale as the uncertainties. A TALYS calculation was performed using default parameters to see how the scale of the data compared and was generally larger than the measured cross sections. For $^{12}$C(n,p$_n$) reactions, TALYS only predicts compound contributions to the cross sections. 

\subsection{\boldmath  $^{12}$C(n,p$_{3+4}$)$^{12}$B}\label{np3+4}
\begin{figure}
\includegraphics[width=\columnwidth]{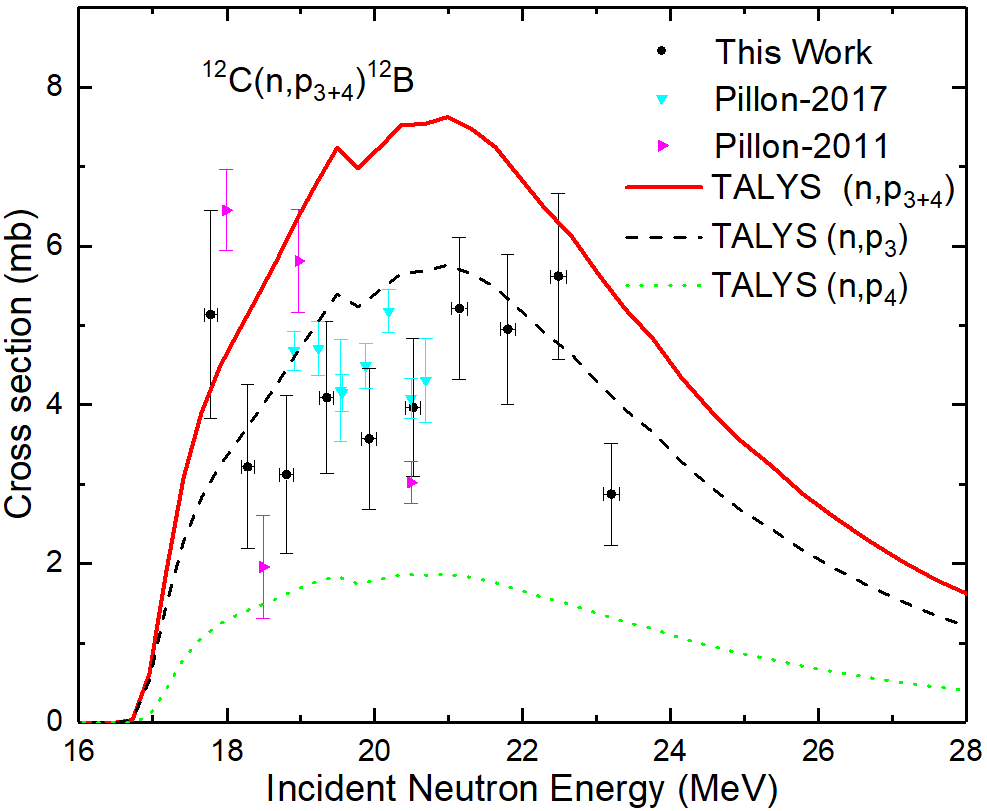}
\caption{\label{fig:np3_xs} Cross sections for $^{12}$C(n,p$_{3+4}$)$^{12}$B, normalized to previous $^{12}$C(n,$\alpha$)$^{9}$Be data. This work and previous data are compared with TALYS calculations performed with default parameters for (n,p$_{3}$), (n,p$_{4}$), and (n,p$_{3+4}$), which results from the sum of (n,p$_{3}$) and (n,p$_{4}$).}
\end{figure}

The similar excitation energies of the third (J$^\pi$ = 1$^-$, E$_x$ = 2.621 MeV) and fourth (J$^\pi$ = 0$^+$, E$_x$ = 2.723 MeV) excited states of $^{12}$B precluded unique identification in this work due to insufficient resolution. Thus, a combined $^{12}$C(n,p$_{3+4}$)$^{12}$B cross section was measured from E$_n =$ 17.78 to 23.2 MeV, as shown in Fig. \ref{fig:np3_xs} and Table \ref{tab:table_p3}. Pillon \textit{et al}. \cite{Pillon2017}, had superior resolution and could partially distinguish the contributions of these two reactions. In our study, a single Gaussian distribution was used to fit the peak resulting from these reaction channels. TALYS calculations using default parameters were performed for (n,p$_3$) and (n,p$_4$) separately, which are shown in addition to the summed (n,p$_{3+4}$) cross sections. While the calculation and data are on the same scale, the scatter of data combined with the general overprediction by TALYS limit their agreement. 

\subsection{\boldmath  $^{12}$C(n,d$_1$)$^{11}$B}\label{nd1}
\begin{figure}
\includegraphics[width=\columnwidth]{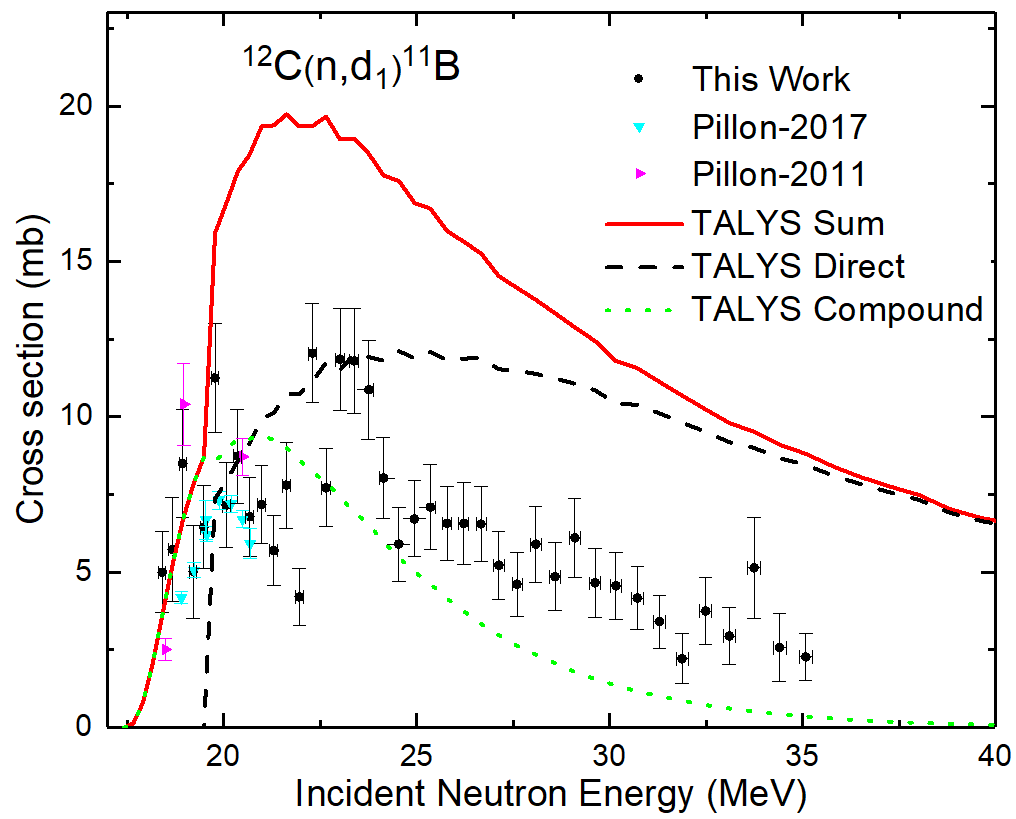}
\caption{\label{fig:nd1_xs} Cross sections for $^{12}$C(n,d$_1$)$^{11}$B, normalized to previous $^{12}$C(n,$\alpha$)$^{9}$Be data. This work and previous data are compared with a TALYS calculation using default parameters. The direct and compound components calculated by TALYS are shown individually, as well as the sum (direct + compound).}
\end{figure}

Cross sections were also determined for production of a deuteron and $^{11}$B in the first excited state (J$^\pi = $ 1/2$^-$, E$_x$ = 2.125 MeV), shown in Fig. \ref{fig:nd1_xs} and Table \ref{tab:table_d1}. The energy range for which cross sections were determined (from E$_n =$ 18.41 to 35 MeV) in this study extends to higher energies than previous studies \cite{Pillon2011, Pillon2017}, which ended at 20.69 MeV. Reasonable agreement is observed between the findings of this work and those of Pillon \textit{et al}. The results are compared to a TALYS calculation using default parameters. The summed (direct + compound) cross section calculated by TALYS is nearly a factor of two higher than the experimental results, driven by the overprediction of the direct component above 25 MeV. This overprediction contrasts with the cross sections calculated by TALYS in Ref. \cite{wantz2025}, which were generally lower than the experimental data. Some structure is evident in the cross section, with a possible resonance observed around 24 MeV incident neutron energy. 

\subsection{\boldmath  $^{12}$C(n,d$_2$)$^{11}$B}\label{nd2}
\begin{figure}
\includegraphics[width=\columnwidth]{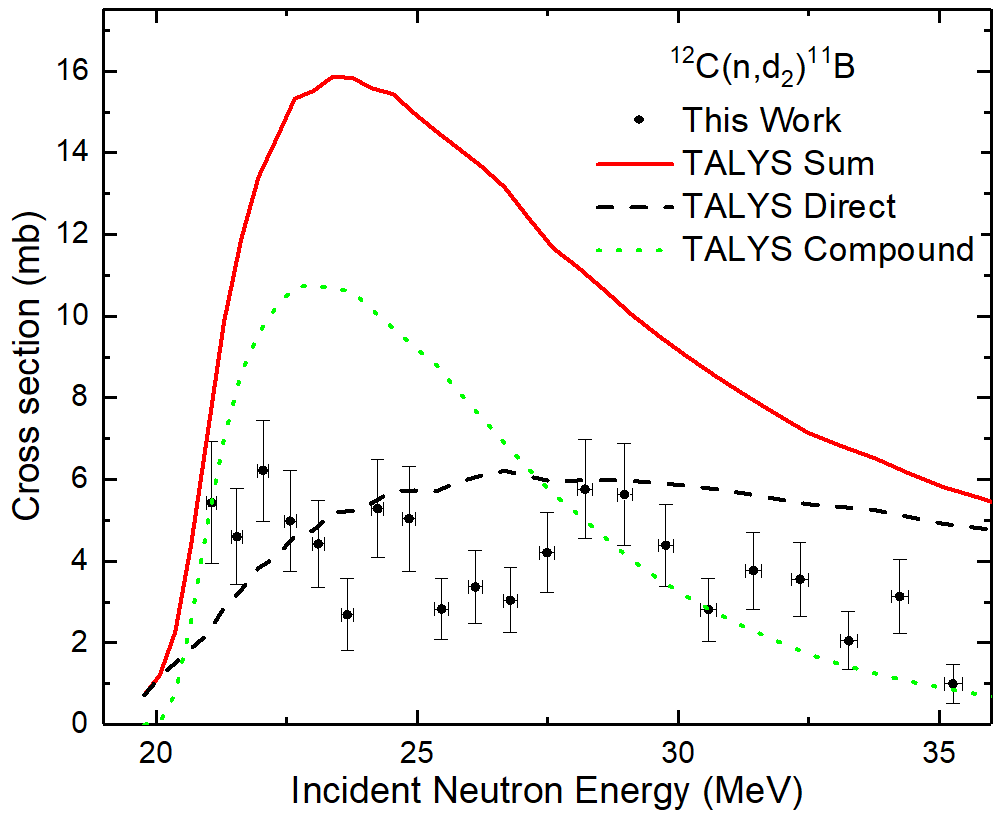}
\caption{\label{fig:nd2_xs} Cross sections for $^{12}$C(n,d$_2$)$^{11}$B, normalized to previous $^{12}$C(n,$\alpha$)$^{9}$Be data. This work is compared with a TALYS calculation using default parameters. The direct and compound components calculated by TALYS are shown individually, as well as the sum (direct + compound).}
\end{figure}

Lastly, cross sections were determined for production of a deuteron and $^{11}$B in the second excited state (J$^\pi = $ 5/2$^-$, E$_x$ = 4.445 MeV), shown in Fig. \ref{fig:nd2_xs} and Table \ref{tab:table_d2}. These are the first measurements of these cross sections, which extend from E$_n =$ 21.06 to 35.26 MeV. Again, the data were compared to a TALYS calculation using default parameters; the overall scale of the calculation is generally more than a factor of two larger than the data. Here, the compound component is overpredicted between 21 and 27 MeV. while the direct component is overpredicted above 29 MeV. Some structure is evident in the cross section, with possible resonances observed around 24.5 and 28.5 MeV. 

\subsection{\boldmath  $^{12}$C(n,p)$^{12}$B total}\label{np_total}
\begin{figure}[H]
\includegraphics[width=\columnwidth]{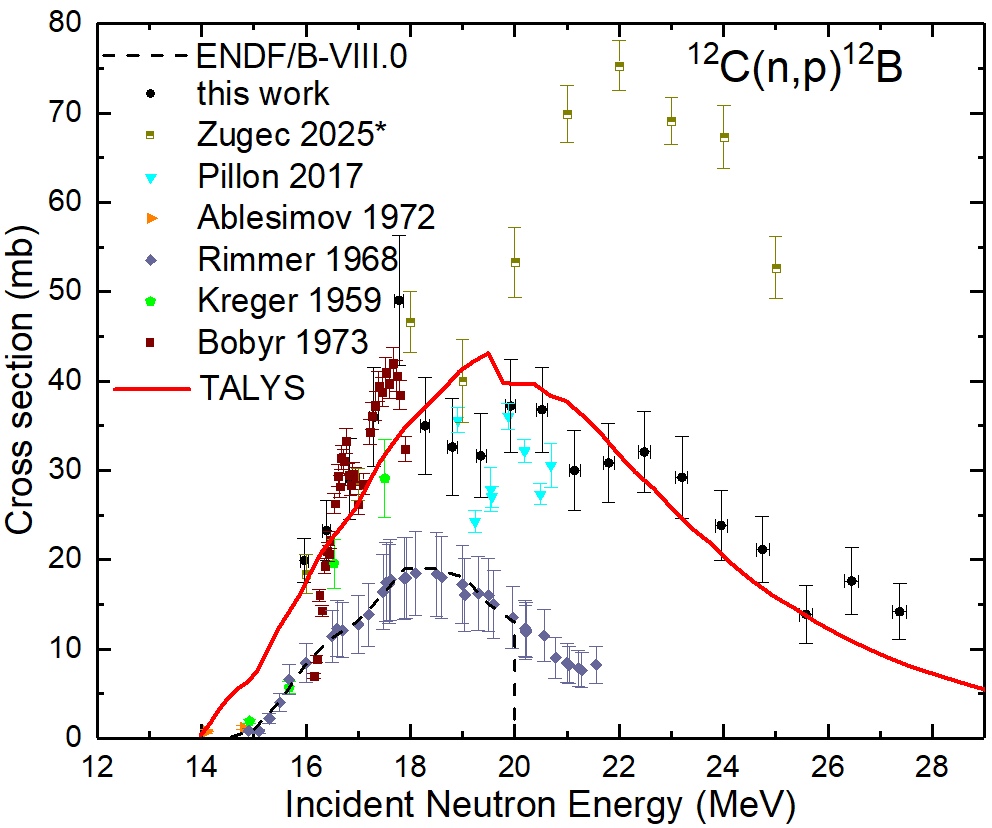}
\caption{\label{fig:np_xs} Cross sections for $^{12}$C(n,p)$^{12}$B. Exclusive cross sections for (n,p$_0$), (n,p$_1$), (n,p$_2$), (n,p$_3$), and (n,p$_4$) were summed for this work and Pillon \textit{et al.} \cite{Pillon2017}, while the earlier studies used the activation technique. Data are compared with the ENDF/B-VIII.0 evaluation and a TALYS calculation using default parameters. *Only the first two data points (16, 17 MeV) for Zugec et al. are $^{12}$C(n,p)$^{12}$B, since they were sensitive to proton production from other reactions above 17.3 MeV. The rest of their data is included for completeness, but is not directly comparable. }
\end{figure}

The total $^{12}$C(n,p)$^{12}$B cross section was extracted by combining the present work with that of Ref. \cite{wantz2025} and  is shown in Fig. \ref{fig:np_xs}. Note that this includes the population of every bound state of $^{12}$B, but not any unbound states. The (n,p) cross sections of the present work, as well as the work of Pillon \textit{et al}. \cite{Pillon2017}, were obtained by summing the contributions of the exclusive cross sections to find the total (n,p) cross section. Since the energy binning varied between the different channels in our study, a linear interpolation was applied to the (n,p$_0$), (n,p$_1$), and (n,p$_2$) channels so that the exclusive cross sections could be summed at the same energies. In general, fairly good agreement is shown with Pillon's 2017 results \cite{Pillon2017}. Pillon's 2011 cross sections \cite{Pillon2011} are not included here, since they could not measure (n,p$_1$). Because (n,p$_1$) contributes significantly to the summed (n,p) cross section, any summed (n,p) cross section that excludes this channel is not directly comparable. 

The data are also compared with older measurements that used the activation technique, measuring the total $^{12}$B production from neutrons on $^{12}$C, including Refs. \cite{RIMMER1968, Ablesimov1971, Kreger1959, Bobyr1973}. Of these, Kreger \cite{Kreger1959} and Bobyr \cite{Bobyr1973} match our data much closer than the Rimmer data \cite{RIMMER1968}. The $^{12}$C(n,p)$^{12}$B$_\text{total}$ ENDF/B-VIII.0 evaluation \cite{BROWN20181} (which is unchanged in ENDF/B-VIII.1 \cite{ENDF/B-VIII.1}) closely follows Rimmer's data, which above 16 MeV is not consistent with more recent measurements. Thus, our new data supports the need to update the ENDF/B- evaluation for $^{12}$C(n,p)$^{12}$B with more recent measurements. Although the default TALYS calculations struggle to accurately reproduce the trends for each partial cross section, the summed (n,p) cross section is in reasonably good agreement with TALYS over the entire energy region covered.

A recent publication \cite{Zugec2025} by the n\_ToF Collaboration (Zugec \textit{et al.}) measured cross sections for $^{12}$C(n,p), which is only directly comparable to the aforementioned studies below 17.3 MeV since they measured the total proton-production cross section for neutrons on $^{12}$C, rather than specifying that a bound state of $^{12}$B was populated. Thus, for incident neutron energies exceeding the threshold for populating unbound states of $^{12}$B (17.3 MeV), the data should not be compared with other studies.

\subsection{\boldmath  $^{12}$C(n,d)$^{11}$B total}\label{nd_total}
\begin{figure}
\includegraphics[width=\columnwidth]{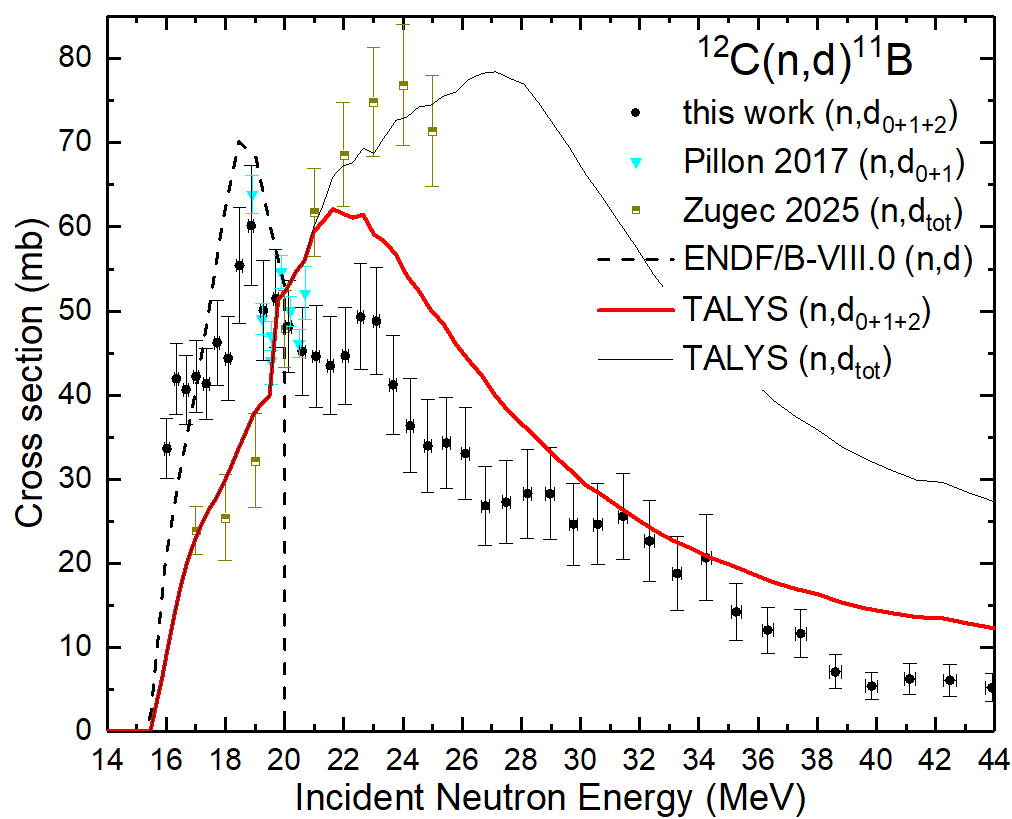}
\caption{\label{fig:nd_xs} Cross sections for $^{12}$C(n,d)$^{11}$B. Exclusive cross sections for (n,d$_0$) and (n,d$_1$) were summed for Pillon \textit{et al.} \cite{Pillon2017}. For this work, (n,d$_0$), (n,d$_1$), and (n,d$_2$) were summed. Zugec \textit{et al.} \cite{Zugec2025} measured total deuteron production, which is only directly comparable to Ref. \cite{Pillon2017} up to 19.3 MeV, when (n,d$_2$) becomes possible, and only directly comparable to the present work up to 21.6 MeV, when (n,d$_3$) becomes possible. Data are compared with the ENDF/B-VIII.0 evaluation and a TALYS calculation using default parameters. TALYS results are included for both (n,d$_{0+1+2}$) and (n,d$_\text{total}$).}
\end{figure}

At low energies, the $^{12}$C(n,d) total cross section is equivalent to the sum of relatively few partial cross sections. Thus, a comparison can be made between this dataset and other studies, which is shown in Fig. \ref{fig:nd_xs}. Similar to the $^{12}$C(n,p)$^{12}$B cross section discussed in the previous section, a linear interpolation of the (n,d$_0$) and (n,d$_1$) cross sections was performed to make the data points align with the coarser-binned (n,d$_2$) data. 

Below 19.3 MeV, the data of this work and Refs. \cite{Pillon2017,Zugec2025} are directly comparable, since the total (n,d) cross section is equivalent to (n,d$_0$) + (n,d$_1$). Above 19.3 MeV, Ref. \cite{Pillon2017} is not directly comparable due to the opening of the (n,d$_2$) channel. Above 21.6 MeV, the present work is also not directly comparable with Ref. \cite{Zugec2025} due to the opening of the (n,d$_3$) channel. Ref. \cite{Pillon2011} is not included here because their (n,d$_0$)  cross section contains contributions from (n,p$_1$). Despite this limited energy range where direct comparisons can be made, it is clear the (n,d$_\text{tot}$) data of Ref. \cite{Zugec2025} differs significantly from the present work and Ref. \cite{Pillon2017}. It is unclear where this discrepancy arises, but the data of Ref. \cite{Zugec2025} is approximately a factor of two lower below 20 MeV. 
While some deviation at higher energies (above 21.6 MeV) is expected since more deuteron-producing channels are above threshold, the discrepancy should not exist at 21 MeV, where their (n,d$_\text{tot}$) cross section is unexpectedly larger. 

The data are also compared with the ENDF/B- evaluation, which was informed by Ref. \cite{PhysRev.106.775}, which measured the inverse reaction. Reasonable agreement is observed between our data and the evaluation below 20 MeV, although the evaluation should be updated to reflect more recent measurements. 

The data are also compared with TALYS calculations performed with default parameters. The TALYS-calculated cross sections for (n,d$_{0+1+2}$) and for $^{12}$C(n,d$_\text{tot}$)$^{11}$B are included. At higher energies, some limited agreement is found between our data and the (n,d$_{0+1+2}$) curve. Good agreement is observed between Ref. \cite{Zugec2025} and the TALYS $^{12}$C(n,d$_\text{tot}$)$^{11}$B cross section.

\section{Conclusions}
Cross sections were determined for four neutron induced reactions on $^{12}$C: $^{12}$C(n,p$_2$)$^{12}$B, $^{12}$C(n,p$_{3+4}$)$^{12}$B, $^{12}$C(n,d$_1$)$^{11}$B, and $^{12}$C(n,d$_2$)$^{11}$B. These measurements will be beneficial as inputs for empirical simulation packages, such as \MCNP, \MENATER, neuSIM4, and some \GEANT $ $ physics lists. Improved cross section data enables enhanced simulation and analysis of detectors that contain carbon nuclei, such as diamond detectors and organic scintillators.

The summed $^{12}$C(n,p)$^{12}$B and $^{12}$C(n,d)$^{11}$B cross sections clearly illustrate the current deficiencies in nuclear data evaluation libraries.  These evaluations were previously derived with sparse experimental constraints and using reaction models that are not necessarily well suited for describing reactions on such light nuclei at such high energies. The measurements presented in this work, including higher energy cross sections for $^{12}$C(n,p$_2$)$^{12}$B, $^{12}$C(n,p$_{3+4}$)$^{12}$B, and $^{12}$C(n,d$_1$)$^{11}$B and the first cross section measurement of $^{12}$C(n,d$_2$)$^{11}$B, supply tighter constraints for reaction models and future nuclear data evaluations.

\begin{acknowledgments}
 This work was supported by the U.S. Department of Energy through the Los Alamos National Laboratory, Contract No. 89233218CNA000001. This work was also supported by the National Science Foundation, grant numbers PHY 2012040, PHY 2310078, PHY 2311125, PHY 2311126, and PHY 2011265, and the U.S. Department of Energy, Office of Science, Office of Nuclear Physics, Award Number DE-SC0022037. 

\end{acknowledgments}

\begin{table*}[h]
\caption{\label{tab:table_p2}
Partial cross sections and uncertainties for $^{12}$C(n,p$_2$)$^{12}$B, normalized to previously measured $^{12}$C(n,$\alpha_0$)$^9$Be experimental data at 14.1 MeV}
\begin{ruledtabular}
\begin{tabular}{c c p{0.0 in} c c p{0.0 in}  c c }
 E$_n$ (MeV) & $\sigma$ (mb) &  & E$_n$ (MeV) & $\sigma$ (mb) & & E$_n$ (MeV) & $\sigma$ (mb) \\ 
 \hline
16.34  &  5.0  $\pm$  1.1  & &  19.28  &  3.2  $\pm$  0.9  & &  22.57  &  2.9  $\pm$  0.7  \\
16.67  &  3.6  $\pm$  1.2  & &  19.71  &  4.9  $\pm$  1.1  & &  23.11  &  3.4  $\pm$  1.0  \\
17.01  &  7.9  $\pm$  1.6  & &  20.14  &  2.8  $\pm$  0.9  & &  23.66  &  2.7  $\pm$  0.7  \\
17.36  &  5.1  $\pm$  1.6  & &  20.60  &  5.1  $\pm$  0.9  & &  24.24  &  1.5  $\pm$  0.6  \\
17.72  &  7.8  $\pm$  1.7  & &  21.06  &  5.3  $\pm$  1.0  & &  24.84  &  2.6  $\pm$  0.8  \\
18.09  &  7.0  $\pm$  1.3  & &  21.55  &  4.1  $\pm$  0.8  & &  25.47  &  2.9  $\pm$  0.9  \\
18.48  &  6.6  $\pm$  1.2  & &  22.05  &  3.5  $\pm$  0.7  & &  26.11  &  1.1  $\pm$  0.7  \\
18.87  &  3.9  $\pm$  1.0  & &  \\
\end{tabular}
\end{ruledtabular}
\end{table*}

\begin{table*}[h]
\caption{\label{tab:table_p3}
Partial cross sections and uncertainties for $^{12}$C(n,p$_{3+4}$)$^{12}$B, normalized to previously measured $^{12}$C(n,$\alpha_0$)$^9$Be experimental data at 14.1 MeV.  The cross sections for (n,p$_3$) and (n,p$_4$) calculated by TALYS are also included.}
\begin{ruledtabular}
\begin{tabular}{c c c c}
 E$_n$  & Measured  & TALYS  & TALYS \\ 
 (MeV) & (n,p$_3$+p$_4$) $\sigma$ (mb) & (n,p$_3$) $\sigma$ (mb) & (n,p$_4$) $\sigma$ (mb) \\
 \hline
17.78  &  5.1  $\pm$  1.0  &  3.0  &  1.2  \\			
18.28  &  3.2  $\pm$  0.8  &  3.7  &  1.4  \\			
18.81  &  3.1  $\pm$  0.8  &  4.4  &  1.6  \\			
19.35  &  4.1  $\pm$  0.8  &  5.2  &  1.8  \\			
19.92  &  4.7  $\pm$  1.1  &  5.3  &  1.8  \\			
20.52  &  4.0  $\pm$  0.8  &  5.7  &  1.9  \\			
21.14  &  5.2  $\pm$  0.9  &  5.7  &  1.8  \\			
21.80  &  5.0  $\pm$  0.9  &  5.3  &  1.7  \\			
22.48  &  5.6  $\pm$  1.0  &  4.8  &  1.5  \\			
23.20  &  2.9  $\pm$  0.6  &  4.1  &  1.3  \\			

\end{tabular}
\end{ruledtabular}
\end{table*}

\begin{table*}[h]
\caption{\label{tab:table_d1}
Partial cross sections and uncertainties for $^{12}$C(n,d$_1$)$^{11}$B, normalized to previously measured $^{12}$C(n,$\alpha_0$)$^9$Be experimental data at 14.1 MeV}
\begin{ruledtabular}
\begin{tabular}{c c p{0.0 in} c c p{0.0 in}  c c }
 E$_n$ (MeV) & $\sigma$ (mb) &  & E$_n$ (MeV) & $\sigma$ (mb) & & E$_n$ (MeV) & $\sigma$ (mb) \\ 
 \hline
18.41  &  5.0  $\pm$  1.3  & &  22.66  &  7.7  $\pm$  1.3  & &  28.09  &  5.9  $\pm$  1.2  \\
18.67  &  5.7  $\pm$  1.7  & &  23.01  &  11.8  $\pm$  1.6  & &  28.59  &  4.9  $\pm$  1.1  \\
18.94  &  8.5  $\pm$  1.7  & &  23.38  &  11.8  $\pm$  1.7  & &  29.10  &  6.1  $\pm$  1.3  \\
19.21  &  5.0  $\pm$  1.5  & &  23.76  &  10.9  $\pm$  1.6  & &  29.62  &  4.7  $\pm$  1.1  \\
19.49  &  6.4  $\pm$  1.3  & &  24.14  &  8.0  $\pm$  1.3  & &  30.16  &  4.6  $\pm$  1.1  \\
19.78  &  11.3  $\pm$  1.7  & &  24.54  &  5.9  $\pm$  1.2  & &  30.72  &  4.2  $\pm$  1.0  \\
20.07  &  7.1  $\pm$  1.4  & &  24.94  &  6.7  $\pm$  1.2  & &  31.29  &  3.4  $\pm$  0.9  \\
20.37  &  8.7  $\pm$  1.5  & &  25.36  &  7.1  $\pm$  1.4  & &  31.88  &  2.2  $\pm$  0.8  \\
20.67  &  6.8  $\pm$  1.3  & &  25.79  &  6.6  $\pm$  1.2  & &  32.48  &  3.7  $\pm$  1.1  \\
20.98  &  7.2  $\pm$  1.3  & &  26.22  &  6.6  $\pm$  1.3  & &  33.10  &  2.9  $\pm$  0.9  \\
21.30  &  5.7  $\pm$  1.1  & &  26.67  &  6.6  $\pm$  1.2  & &  33.75  &  5.1  $\pm$  1.6  \\
21.63  &  7.8  $\pm$  1.4  & &  27.13  &  5.2  $\pm$  1.1  & &  34.41  &  2.6  $\pm$  1.1  \\
21.96  &  4.2  $\pm$  0.9  & &  27.60  &  4.6  $\pm$  1.0  & &  35.08  &  2.3  $\pm$  0.8  \\
22.31  &  12.0  $\pm$  1.6 & & & & & & \\
\end{tabular}
\end{ruledtabular}
\end{table*}

\begin{table*}[h]
\caption{\label{tab:table_d2}
Partial cross sections and uncertainties for $^{12}$C(n,d$_2$)$^{11}$B, normalized to previously measured $^{12}$C(n,$\alpha_0$)$^9$Be experimental data at 14.1 MeV}
\begin{ruledtabular}
\begin{tabular}{c c p{0.0 in} c c p{0.0 in}  c c }
 E$_n$ (MeV) & $\sigma$ (mb) &  & E$_n$ (MeV) & $\sigma$ (mb) & & E$_n$ (MeV) & $\sigma$ (mb) \\ 
 \hline
21.06  &  5.4  $\pm$  1.5  & &  24.84  &  5.0  $\pm$  1.3  & &  29.76  &  4.4  $\pm$  1.0  \\
21.55  &  4.6  $\pm$  1.2  & &  25.47  &  2.8  $\pm$  0.7  & &  30.58  &  2.8  $\pm$  0.8  \\
22.05  &  6.2  $\pm$  1.2  & &  26.11  &  3.4  $\pm$  0.9  & &  31.44  &  3.8  $\pm$  0.9  \\
22.57  &  5.0  $\pm$  1.2  & &  26.79  &  3.0  $\pm$  0.8  & &  32.33  &  3.6  $\pm$  0.9  \\
23.11  &  4.4  $\pm$  1.1  & &  27.49  &  4.2  $\pm$  1.0  & &  33.27  &  2.1  $\pm$  0.7  \\
23.66  &  2.7  $\pm$  0.9  & &  28.21  &  5.8  $\pm$  1.2  & &  34.24  &  3.1  $\pm$  0.9  \\
24.24  &  5.3  $\pm$  1.2  & &  28.97  &  5.6  $\pm$  1.3  & &  35.26  &  1.0  $\pm$  0.5  \\
\end{tabular}
\end{ruledtabular}
\end{table*}


\providecommand{\noopsort}[1]{}\providecommand{\singleletter}[1]{#1}%

\end{document}


\maketitle

\section{Data tables for each reaction channel}
\LTcapwidth=0.9\textwidth

\begin{longtable}[h]{|c|c|c|c|c|c|c|c|c|c|c|c|c|c|c|c|c|}
 \caption{$^{12}$C(n,d$_1$)$^{11}$B supplemental table. The table contains the incident neutron energy ($E_n$); counts from $^{12}$C(n,d$_1$)$^{11}$B (Counts: Peak); total counts in the fitting region (peak and background, Counts: Total); the charged particle detection efficiency correction for this channel (Corrections: Efficiency); the simulated neutron transmission correction for the flux distribution (Corrections: Transmission); the cross section for this channel (Cross section: $\sigma$ (mb)); the statistical uncertainties, broken into separate contributions from the counts from (n,d$_1$) at each energy (Statistical Uncertainty: $\sigma_{Peak}$), from the the background counts at each energy (Statistical Uncertainty: $\sigma_{Bkg}$), and from the (n,$\alpha$) counts at 14.1 MeV used for normalization (Statistical uncertainties: $\sigma_{N_{n,\alpha}}$); the total statistical uncertainties combining all statistical uncertainties (Statistical Uncertainties: $\sigma_{stat. total}$); and systematic uncertainties, which include the charged particle detection efficiency correction uncertainty (Systematic Uncertainties: $\sigma_{Eff.}$), the relative flux distribution uncertainty (Systematic Uncertainties:  $\sigma_{Flux}$), the fit/integration uncertainties (Systematic Uncertainties: $\sigma_{Fit}$), the incident neutron energy reconstruction uncertainty (systematic uncertainties: $\sigma_{NKE}$), and the normalization uncertainty based on the weighted average uncertainty of previous data (Systematic Uncertainties: $\sigma_{Norm.}$). The relative uncertainty, which includes both statistical and systematic uncertainties, is included as a percentage, and the absolute uncertainty is included in millibarns. \label{n,d1}}\\
 \hline
$E_n$  & \multicolumn{2}{c|}{Counts} & \multicolumn{2}{c|}{Corrections (\%)} & Cross section & \multicolumn{4}{c|}{Statistical Uncertainties (\%)} &  \multicolumn{5}{c|}{Systematic Uncertainties (\%)} & Rel. Unc.(\%) & Abs. Unc.\\
\hline    
(MeV) & Peak & Total & Efficiency & Transmission & $\sigma$ (mb) & $\sigma_{Peak}$ &  $\sigma_{Bkg}$ &  $\sigma_{N_{n,\alpha}}$ & $\sigma_{stat. total}$ & $\sigma_{Eff.}$ & $\sigma_{Flux}$ & $\sigma_{Fit}$ & $\sigma_{NKE}$ & $\sigma_{Norm.}$ & $\sigma_{Rel.}$ & (mb)\\
 \hline
 \endfirsthead
 \hline
$E_n$  & \multicolumn{2}{c|}{Counts} & \multicolumn{2}{c|}{Corrections (\%)} & Cross section & \multicolumn{4}{c|}{Statistical Uncertainties (\%)} &  \multicolumn{5}{c|}{Systematic Uncertainties (\%)} & Rel. Unc.(\%) & Abs. Unc.\\ \hline
(MeV) & Peak & Total & Efficiency & Transmission & $\sigma$ (mb) & $\sigma_{Peak}$ &  $\sigma_{Bkg}$ &  $\sigma_{N_{n,\alpha}}$ & $\sigma_{stat. total}$ & $\sigma_{Eff.}$ & $\sigma_{Flux}$ & $\sigma_{Fit}$ & $\sigma_{NKE}$ & $\sigma_{Norm.}$ & $\sigma_{Rel.}$ &  (mb)\\ \hline
 \endhead

 \hline
 \endfoot
18.41  &  87  &  351  &  99.6  &  81.6  &  5.0  &  10.7  &  18.8  &  2.96  &  21.8  &  4  &  7.3  &  10.6  &  0.49  &  4.87  &  26.1  &  1.3  \\
18.67  &  100  &  547  &  99.6  &  81.7  &  5.7  &  10.0  &  21.2  &  2.96  &  23.7  &  4  &  6.8  &  14.6  &  0.49  &  4.87  &  29.3  &  1.7  \\
18.94  &  147  &  587  &  99.5  &  81.7  &  8.5  &  8.2  &  14.2  &  2.96  &  16.7  &  4  &  6.3  &  7.9  &  0.49  &  4.87  &  20.5  &  1.7  \\
19.21  &  86  &  443  &  99.4  &  81.1  &  5.0  &  10.8  &  21.9  &  2.96  &  24.6  &  4  &  5.6  &  14.5  &  0.49  &  4.87  &  29.8  &  1.5  \\
19.49  &  111  &  376  &  99.0  &  80.9  &  6.4  &  9.5  &  14.7  &  2.96  &  17.8  &  4  &  5.5  &  6.9  &  0.49  &  4.87  &  20.8  &  1.3  \\
19.78  &  195  &  501  &  98.8  &  80.9  &  11.3  &  7.2  &  9.0  &  2.96  &  11.9  &  4  &  5.5  &  5.6  &  0.49  &  4.87  &  15.5  &  1.7  \\
20.07  &  125  &  347  &  98.5  &  81.1  &  7.1  &  9.0  &  11.9  &  2.96  &  15.2  &  4  &  5.4  &  8.1  &  0.49  &  4.87  &  19.1  &  1.4  \\
20.37  &  154  &  402  &  98.1  &  81.5  &  8.7  &  8.1  &  10.2  &  2.96  &  13.3  &  4  &  5.5  &  7.1  &  0.49  &  4.87  &  17.3  &  1.5  \\
20.67  &  122  &  337  &  97.7  &  81.9  &  6.8  &  9.1  &  12.0  &  2.96  &  15.3  &  4  &  5.8  &  6.3  &  0.49  &  4.87  &  18.7  &  1.3  \\
20.98  &  128  &  299  &  97.2  &  81.9  &  7.2  &  8.8  &  10.2  &  2.96  &  13.8  &  4  &  5.3  &  6.7  &  0.49  &  4.87  &  17.4  &  1.3  \\
21.30  &  103  &  242  &  96.8  &  82.1  &  5.7  &  9.9  &  11.5  &  2.96  &  15.4  &  4  &  5.4  &  8.7  &  0.49  &  4.87  &  19.6  &  1.1  \\
21.63  &  143  &  340  &  96.3  &  82.2  &  7.8  &  8.4  &  9.8  &  2.96  &  13.3  &  4  &  6.0  &  8.2  &  0.49  &  4.87  &  17.8  &  1.4  \\
21.96  &  77  &  204  &  95.6  &  82.3  &  4.2  &  11.4  &  14.6  &  2.96  &  18.7  &  4  &  6.1  &  6.2  &  0.49  &  4.87  &  21.6  &  0.9  \\
22.31  &  222  &  385  &  95.3  &  82.1  &  12.0  &  6.7  &  5.8  &  2.96  &  9.3  &  4  &  6.0  &  3.6  &  0.49  &  4.87  &  13.3  &  1.6  \\
22.66  &  142  &  289  &  94.4  &  82.2  &  7.7  &  8.4  &  8.6  &  2.96  &  12.4  &  4  &  6.0  &  6.1  &  0.49  &  4.87  &  16.3  &  1.3  \\
23.01  &  220  &  408  &  93.9  &  82.0  &  11.8  &  6.7  &  6.2  &  2.96  &  9.7  &  4  &  6.5  &  4.0  &  0.49  &  4.87  &  13.8  &  1.6  \\
23.38  &  224  &  399  &  93.5  &  82.1  &  11.8  &  6.7  &  5.9  &  2.96  &  9.4  &  4  &  7.4  &  4.3  &  0.49  &  4.87  &  14.2  &  1.7  \\
23.76  &  210  &  339  &  92.8  &  81.9  &  10.9  &  6.9  &  5.4  &  2.96  &  9.2  &  4  &  8.5  &  4.4  &  0.49  &  4.87  &  14.7  &  1.6  \\
24.14  &  157  &  288  &  92.0  &  82.0  &  8.0  &  8.0  &  7.3  &  2.96  &  11.2  &  4  &  9.1  &  4.0  &  0.49  &  4.87  &  16.2  &  1.3  \\
24.54  &  115  &  309  &  90.9  &  82.1  &  5.9  &  9.3  &  12.1  &  2.96  &  15.5  &  4  &  9.3  &  6.6  &  0.49  &  4.87  &  20.2  &  1.2  \\
24.94  &  131  &  287  &  90.0  &  82.2  &  6.7  &  8.7  &  9.5  &  2.96  &  13.3  &  4  &  9.3  &  5.3  &  0.49  &  4.87  &  18.2  &  1.2  \\
25.36  &  140  &  339  &  89.3  &  82.5  &  7.1  &  8.5  &  10.1  &  2.96  &  13.5  &  4  &  9.7  &  7.7  &  0.49  &  4.87  &  19.4  &  1.4  \\
25.79  &  129  &  271  &  88.2  &  82.6  &  6.6  &  8.8  &  9.2  &  2.96  &  13.1  &  4  &  10.0  &  3.9  &  0.49  &  4.87  &  18.0  &  1.2  \\
26.22  &  128  &  304  &  87.0  &  82.6  &  6.6  &  8.8  &  10.3  &  2.96  &  13.9  &  4  &  10.2  &  8.5  &  0.49  &  4.87  &  20.2  &  1.3  \\
26.67  &  129  &  249  &  86.1  &  83.1  &  6.6  &  8.8  &  8.5  &  2.96  &  12.6  &  4  &  10.8  &  5.6  &  0.49  &  4.87  &  18.6  &  1.2  \\
27.13  &  105  &  237  &  84.4  &  83.3  &  5.2  &  9.8  &  11.0  &  2.96  &  15.0  &  4  &  12.3  &  4.4  &  0.49  &  4.87  &  20.9  &  1.1  \\
27.60  &  91  &  212  &  83.5  &  83.4  &  4.6  &  10.5  &  12.0  &  2.96  &  16.2  &  4  &  12.3  &  6.5  &  0.49  &  4.87  &  22.2  &  1.0  \\
28.09  &  114  &  240  &  81.5  &  83.3  &  5.9  &  9.3  &  9.8  &  2.96  &  13.9  &  4  &  12.5  &  6.4  &  0.49  &  4.87  &  20.7  &  1.2  \\
28.59  &  96  &  219  &  80.1  &  83.6  &  4.9  &  10.2  &  11.6  &  2.96  &  15.7  &  4  &  13.8  &  5.4  &  0.49  &  4.87  &  22.5  &  1.1  \\
29.10  &  121  &  231  &  78.8  &  83.6  &  6.1  &  9.1  &  8.7  &  2.96  &  12.9  &  4  &  14.0  &  5.5  &  0.49  &  4.87  &  20.8  &  1.3  \\
29.62  &  91  &  221  &  76.6  &  83.7  &  4.7  &  10.5  &  12.6  &  2.96  &  16.7  &  4  &  14.0  &  7.0  &  0.49  &  4.87  &  23.7  &  1.1  \\
30.16  &  87  &  194  &  74.7  &  83.7  &  4.6  &  10.7  &  11.8  &  2.96  &  16.2  &  4  &  13.8  &  8.1  &  0.49  &  4.87  &  23.6  &  1.1  \\
30.72  &  79  &  175  &  72.5  &  84.0  &  4.2  &  11.3  &  12.5  &  2.96  &  17.1  &  4  &  13.6  &  7.9  &  0.49  &  4.87  &  24.1  &  1.0  \\
31.29  &  64  &  144  &  70.3  &  84.2  &  3.4  &  12.5  &  14.0  &  2.96  &  19.0  &  4  &  13.9  &  5.4  &  0.49  &  4.87  &  25.0  &  0.9  \\
31.88  &  41  &  144  &  67.9  &  84.3  &  2.2  &  15.7  &  25.0  &  2.96  &  29.6  &  4  &  14.1  &  13.4  &  0.49  &  4.87  &  36.0  &  0.8  \\
32.48  &  68  &  198  &  65.4  &  84.3  &  3.7  &  12.1  &  16.7  &  2.96  &  20.8  &  4  &  14.9  &  10.8  &  0.49  &  4.87  &  28.5  &  1.1  \\
33.10  &  52  &  166  &  62.4  &  84.6  &  2.9  &  13.8  &  20.4  &  2.96  &  24.8  &  4  &  15.3  &  8.7  &  0.49  &  4.87  &  31.1  &  0.9  \\
33.75  &  89  &  283  &  59.4  &  84.7  &  5.1  &  10.6  &  15.7  &  2.96  &  19.2  &  4  &  15.6  &  18.7  &  0.49  &  4.87  &  31.7  &  1.6  \\
34.41  &  44  &  189  &  56.3  &  85.0  &  2.6  &  15.2  &  27.7  &  2.96  &  31.7  &  4  &  16.5  &  22.7  &  0.49  &  4.87  &  42.8  &  1.1  \\
35.08  &  37  &  88  &  53.1  &  84.9  &  2.3  &  16.4  &  19.1  &  2.96  &  25.3  &  4  &  17.5  &  10.3  &  0.49  &  4.87  &  33.1  &  0.8  \\

\hline 
 \end{longtable}

\begin{longtable}[h]{|c|c|c|c|c|c|c|c|c|c|c|c|c|c|c|c|c|}
 \caption{$^{12}$C(n,d$_2$)$^{11}$B supplemental table. The table contains the incident neutron energy ($E_n$); counts from $^{12}$C(n,d$_2$)$^{11}$B (Counts: Peak); total counts in the fitting region (peak and background, Counts: Total); the charged particle detection efficiency correction for this channel (Corrections: Efficiency); the simulated neutron transmission correction for the flux distribution (Corrections: Transmission); the cross section for this channel (Cross section: $\sigma$ (mb)); the statistical uncertainties, broken into separate contributions from the counts from (n,d$_2$) at each energy (Statistical Uncertainty: $\sigma_{Peak}$), from the the background counts at each energy (Statistical Uncertainty: $\sigma_{Bkg}$), and from the (n,$\alpha$) counts at 14.1 MeV used for normalization (Statistical uncertainties: $\sigma_{N_{n,\alpha}}$); the total statistical uncertainties combining all statistical uncertainties (Statistical Uncertainties: $\sigma_{stat. total}$); and systematic uncertainties, which include the charged particle detection efficiency correction uncertainty (Systematic Uncertainties: $\sigma_{Eff.}$), the relative flux distribution uncertainty (Systematic Uncertainties:  $\sigma_{Flux}$), the fit/integration uncertainties (Systematic Uncertainties: $\sigma_{Fit}$), the incident neutron energy reconstruction uncertainty (systematic uncertainties: $\sigma_{NKE}$), and the normalization uncertainty based on the weighted average uncertainty of previous data (Systematic Uncertainties: $\sigma_{Norm.}$). The relative uncertainty, which includes both statistical and systematic uncertainties, is included as a percentage, and the absolute uncertainty is included in millibarns. \label{n,d2}}\\
 \hline
$E_n$  & \multicolumn{2}{c|}{Counts} & \multicolumn{2}{c|}{Corrections (\%)} & Cross section & \multicolumn{4}{c|}{Statistical Uncertainties (\%)} &  \multicolumn{5}{c|}{Systematic Uncertainties (\%)} & Rel. Unc.(\%) & Abs. Unc.\\
\hline    
(MeV) & Peak & Total & Efficiency & Transmission & $\sigma$ (mb) & $\sigma_{Peak}$ &  $\sigma_{Bkg}$ &  $\sigma_{N_{n,\alpha}}$ & $\sigma_{stat. total}$ & $\sigma_{Eff.}$ & $\sigma_{Flux}$ & $\sigma_{Fit}$ & $\sigma_{NKE}$ & $\sigma_{Norm.}$ & $\sigma_{Rel.}$ &  (mb)\\
 \hline
 \endfirsthead
 \hline
$E_n$  & \multicolumn{2}{c|}{Counts} & \multicolumn{2}{c|}{Corrections (\%)} & Cross section & \multicolumn{4}{c|}{Statistical Uncertainties (\%)} &  \multicolumn{5}{c|}{Systematic Uncertainties (\%)} & Rel. Unc.(\%) & Abs. Unc.\\ \hline
(MeV) & Peak & Total & Efficiency & Transmission & $\sigma$ (mb) & $\sigma_{Peak}$ &  $\sigma_{Bkg}$ &  $\sigma_{N_{n,\alpha}}$ & $\sigma_{stat. total}$ & $\sigma_{Eff.}$ & $\sigma_{Flux}$ & $\sigma_{Fit}$ & $\sigma_{NKE}$ & $\sigma_{Norm.}$ & $\sigma_{Rel.}$ &  (mb)\\ \hline
 \endhead

 \hline
 \endfoot

21.06  &  150  &  547  &  99.6  &  82.0  &  5.4  &  8.2  &  22.9  &  2.96  &  24.5  &  4  &  5.2  &  9.6  &  0.49  &  4.87  &  27.5  &  1.5  \\
21.55  &  130  &  587  &  99.4  &  82.0  &  4.6  &  8.8  &  21.1  &  2.96  &  23.1  &  4  &  5.9  &  7.1  &  0.49  &  4.87  &  25.6  &  1.2  \\
22.05  &  179  &  443  &  99.1  &  82.4  &  6.2  &  7.5  &  14.2  &  2.96  &  16.3  &  4  &  6.1  &  7.1  &  0.49  &  4.87  &  19.9  &  1.2  \\
22.57  &  144  &  376  &  98.6  &  82.0  &  5.0  &  8.3  &  20.0  &  2.96  &  21.9  &  4  &  5.9  &  8.0  &  0.49  &  4.87  &  24.8  &  1.2  \\
23.11  &  130  &  501  &  98.0  &  82.1  &  4.4  &  8.8  &  17.8  &  2.96  &  20.1  &  4  &  6.7  &  9.3  &  0.49  &  4.87  &  24.0  &  1.1  \\
23.66  &  82  &  347  &  97.2  &  82.0  &  2.7  &  11.1  &  27.0  &  2.96  &  29.3  &  4  &  8.3  &  11.1  &  0.49  &  4.87  &  33.0  &  0.9  \\
24.24  &  164  &  402  &  96.3  &  81.9  &  5.3  &  7.8  &  15.3  &  2.96  &  17.4  &  4  &  9.3  &  9.7  &  0.49  &  4.87  &  22.8  &  1.2  \\
24.84  &  157  &  337  &  95.4  &  82.5  &  5.0  &  8.0  &  18.5  &  2.96  &  20.4  &  4  &  9.3  &  10.3  &  0.49  &  4.87  &  25.4  &  1.3  \\
25.47  &  89  &  299  &  94.4  &  82.4  &  2.8  &  10.6  &  18.1  &  2.96  &  21.2  &  4  &  9.8  &  10.4  &  0.49  &  4.87  &  26.3  &  0.7  \\
26.11  &  106  &  242  &  93.1  &  82.6  &  3.4  &  9.7  &  19.1  &  2.96  &  21.6  &  4  &  10.1  &  10  &  0.49  &  4.87  &  26.6  &  0.9  \\
26.79  &  97  &  340  &  91.7  &  83.2  &  3.0  &  10.1  &  18.4  &  2.96  &  21.2  &  4  &  11.1  &  7.6  &  0.49  &  4.87  &  25.9  &  0.8  \\
27.49  &  136  &  204  &  90.2  &  83.4  &  4.2  &  8.6  &  14.3  &  2.96  &  16.9  &  4  &  12.4  &  8.2  &  0.49  &  4.87  &  23.4  &  1.0  \\
28.21  &  185  &  385  &  88.5  &  83.3  &  5.8  &  7.4  &  11.0  &  2.96  &  13.6  &  4  &  13.0  &  7.3  &  0.49  &  4.87  &  21.1  &  1.2  \\
28.97  &  184  &  289  &  86.6  &  83.6  &  5.6  &  7.4  &  12.0  &  2.96  &  14.4  &  4  &  14.0  &  7.1  &  0.49  &  4.87  &  22.2  &  1.3  \\
29.76  &  142  &  408  &  84.3  &  83.7  &  4.4  &  8.4  &  13.9  &  2.96  &  16.5  &  4  &  13.9  &  4.5  &  0.49  &  4.87  &  23.0  &  1.0  \\
30.58  &  90  &  399  &  81.7  &  84.1  &  2.8  &  10.5  &  17.6  &  2.96  &  20.7  &  4  &  13.6  &  9.3  &  0.49  &  4.87  &  27.2  &  0.8  \\
31.44  &  120  &  339  &  78.8  &  84.2  &  3.8  &  9.1  &  14.6  &  2.96  &  17.5  &  4  &  14.0  &  9.1  &  0.49  &  4.87  &  25.0  &  0.9  \\
32.33  &  113  &  288  &  76.1  &  84.3  &  3.6  &  9.4  &  14.7  &  2.96  &  17.7  &  4  &  14.7  &  9.0  &  0.49  &  4.87  &  25.5  &  0.9  \\
33.27  &  64  &  309  &  72.4  &  84.6  &  2.1  &  12.5  &  25.5  &  2.96  &  28.6  &  4  &  15.3  &  11.3  &  0.49  &  4.87  &  34.9  &  0.7  \\
34.24  &  97  &  287  &  68.6  &  84.9  &  3.1  &  10.2  &  17.1  &  2.96  &  20.1  &  4  &  16.3  &  10.4  &  0.49  &  4.87  &  28.6  &  0.9  \\
35.26  &  30  &  339  &  64.0  &  84.9  &  1.0  &  18.3  &  39.3  &  2.96  &  43.5  &  4  &  17.7  &  10.8  &  0.49  &  4.87  &  48.6  &  0.5  \\

\hline 
 \end{longtable}
 
\captionsetup{width=.9\textwidth}
\begin{table}[]
\caption{\label{tab:n_p2_xs} $^{12}$C(n,p$_2$)$^{12}$B supplemental table. The table contains the incident neutron energy ($E_n$); counts from $^{12}$C(n,p$_2$)$^{12}$B (Counts: Peak); total counts in the fitting region (peak and background, Counts: Total); the charged particle detection efficiency correction for this channel (Corrections: Efficiency); the simulated neutron transmission correction for the flux distribution (Corrections: Transmission); the cross section for this channel (Cross section: $\sigma$ (mb)); the statistical uncertainties, broken into separate contributions from the counts from (n,p$_2$) at each energy (Statistical Uncertainty: $\sigma_{Peak}$), from the the background counts at each energy (Statistical Uncertainty: $\sigma_{Bkg}$), and from the (n,$\alpha$) counts at 14.1 MeV used for normalization (Statistical uncertainties: $\sigma_{N_{n,\alpha}}$); the total statistical uncertainties combining all statistical uncertainties (Statistical Uncertainties: $\sigma_{stat. total}$); and systematic uncertainties, which include the charged particle detection efficiency correction uncertainty (Systematic Uncertainties: $\sigma_{Eff.}$), the relative flux distribution uncertainty (Systematic Uncertainties:  $\sigma_{Flux}$), the fit/integration uncertainties (Systematic Uncertainties: $\sigma_{Fit}$), the incident neutron energy reconstruction uncertainty (systematic uncertainties: $\sigma_{NKE}$), and the normalization uncertainty based on the weighted average uncertainty of previous data (Systematic Uncertainties: $\sigma_{Norm.}$). The relative uncertainty, which includes both statistical and systematic uncertainties, is included as a percentage, and the absolute uncertainty is included in millibarns.}
    \centering
    \begin{tabular}{|c|c|c|c|c|c|c|c|c|c|c|c|c|c|c|c|c|}
    \hline
$E_n$ & \multicolumn{2}{c|}{Counts} & \multicolumn{2}{c|}{Corrections (\%) } & Cross section & \multicolumn{4}{c|}{Statistical Uncertainties (\%)} &  \multicolumn{5}{c|}{Systematic Uncertainties (\%)} & Rel. Unc.(\%) & Abs. Unc.\\
\hline    
(MeV) & Peak & Total & Efficiency & Transmission & $\sigma$ (mb) & $\sigma_{Peak}$ &  $\sigma_{Bkg}$ &  $\sigma_{N_{n,\alpha}}$ & $\sigma_{stat. total}$ & $\sigma_{Eff.}$ & $\sigma_{Flux}$ & $\sigma_{Fit}$ & $\sigma_{NKE}$ & $\sigma_{Norm.}$ & $\sigma_{Rel.}$ &  (mb)\\
\hline
16.34 & 122 & 570 & 98.7 & 81.2 & 5.0 & 9.0 & 17.3 & 2.96 & 19.7 & 4 & 5.0 & 5.5 & 0.49 & 4.87 & 22.0 & 1.1 \\
16.67 & 90 & 644 & 98.1 & 81.7 & 3.6 & 10.6 & 26.3 & 2.96 & 28.5 & 4 & 5.0 & 15.0 & 0.49 & 4.87 & 33.2 & 1.2 \\
17.01 & 197 & 1091 & 97.7 & 82.0 & 7.9 & 7.1 & 15.2 & 2.96 & 17.1 & 4 & 5.0 & 7.6 & 0.49 & 4.87 & 20.3 & 1.6 \\
17.36 & 128 & 995 & 97.6 & 82.3 & 5.1 & 8.9 & 23.1 & 2.96 & 24.9 & 4 & 5.3 & 17.5 & 0.49 & 4.87 & 31.5 & 1.6 \\
17.72 & 196 & 766 & 97.2 & 82.1 & 7.8 & 7.1 & 12.2 & 2.96 & 14.4 & 4 & 6.9 & 13.0 & 0.49 & 4.87 & 21.6 & 1.7 \\
18.09 & 176 & 646 & 95.9 & 81.8 & 7.0 & 7.5 & 12.3 & 2.96 & 14.8 & 4 & 7.5 & 6.1 & 0.49 & 4.87 & 18.7 & 1.3 \\
18.48 & 166 & 490 & 94.8 & 81.7 & 6.6 & 7.8 & 10.9 & 2.96 & 13.7 & 4 & 7.2 & 5.5 & 0.49 & 4.87 & 17.6 & 1.2 \\
18.87 & 96 & 441 & 93.7 & 81.8 & 3.9 & 10.2 & 19.4 & 2.96 & 22.1 & 4 & 6.4 & 8.3 & 0.49 & 4.87 & 25.3 & 1.0 \\
19.28 & 77 & 324 & 92.4 & 81.1 & 3.2 & 11.4 & 20.5 & 2.96 & 23.6 & 4 & 5.5 & 9.1 & 0.49 & 4.87 & 26.7 & 0.9 \\
19.71 & 118 & 466 & 91.6 & 80.7 & 4.9 & 9.2 & 15.9 & 2.96 & 18.6 & 4 & 5.5 & 8.0 & 0.49 & 4.87 & 21.9 & 1.1 \\
20.14 & 67 & 297 & 89.9 & 81.2 & 2.8 & 12.2 & 22.6 & 2.96 & 25.9 & 4 & 5.4 & 15.5 & 0.49 & 4.87 & 31.3 & 0.9 \\
20.6 & 125 & 299 & 88.3 & 81.8 & 5.1 & 9.0 & 10.6 & 2.96 & 14.2 & 4 & 5.8 & 5.3 & 0.49 & 4.87 & 17.4 & 0.9 \\
21.06 & 129 & 324 & 86.9 & 82.0 & 5.3 & 8.8 & 10.8 & 2.96 & 14.3 & 4 & 5.2 & 9.2 & 0.49 & 4.87 & 18.9 & 1.0 \\
21.55 & 99 & 239 & 84.6 & 82.0 & 4.1 & 10.0 & 11.9 & 2.96 & 15.9 & 4 & 5.9 & 7.5 & 0.49 & 4.87 & 19.6 & 0.8 \\
22.05 & 84 & 215 & 83.1 & 82.4 & 3.5 & 10.9 & 13.6 & 2.96 & 17.7 & 4 & 6.1 & 7.2 & 0.49 & 4.87 & 21.1 & 0.7 \\
22.57 & 68 & 216 & 79.8 & 82.0 & 2.9 & 12.1 & 17.8 & 2.96 & 21.7 & 4 & 5.9 & 6.9 & 0.49 & 4.87 & 24.4 & 0.7 \\
23.11 & 81 & 368 & 78.2 & 82.1 & 3.4 & 11.1 & 21.1 & 2.96 & 24.0 & 4 & 6.7 & 12.0 & 0.49 & 4.87 & 28.4 & 1.0 \\
23.66 & 64 & 201 & 75.8 & 82.0 & 2.7 & 12.5 & 18.4 & 2.96 & 22.5 & 4 & 8.3 & 9.8 & 0.49 & 4.87 & 26.6 & 0.7 \\
24.24 & 36 & 144 & 72.6 & 81.9 & 1.5 & 16.7 & 28.9 & 2.96 & 33.5 & 4 & 9.3 & 14.9 & 0.49 & 4.87 & 38.4 & 0.6 \\
24.84 & 58 & 222 & 68.6 & 82.5 & 2.6 & 13.1 & 21.9 & 2.96 & 25.6 & 4 & 9.3 & 9.9 & 0.49 & 4.87 & 29.7 & 0.8 \\
25.47 & 61 & 251 & 64.5 & 82.4 & 2.9 & 12.8 & 22.4 & 2.96 & 26.0 & 4 & 9.8 & 12.7 & 0.49 & 4.87 & 31.2 & 0.9 \\
26.11 & 24 & 157 & 61.0 & 82.6 & 1.1 & 20.5 & 48.7 & 2.96 & 53.0 & 4 & 10.1 & 28.9 & 0.49 & 4.87 & 61.5 & 0.7 \\

\hline
    \end{tabular}
\end{table}

\begin{table}[]
\caption{\label{tab:n_p3+p4_xs} $^{12}$C(n,p$_{3+4}$)$^{12}$B supplemental table. The table contains the incident neutron energy ($E_n$); counts from $^{12}$C(n,p$_{3+4}$)$^{12}$B (Counts: Peak); total counts in the fitting region (peak and background, Counts: Total); the charged particle detection efficiency correction for this channel (Corrections: Efficiency); the simulated neutron transmission correction for the flux distribution (Corrections: Transmission); the cross section for this channel (Cross section: $\sigma$ (mb)); the statistical uncertainties, broken into separate contributions from the counts from (n,p$_{3+4}$) at each energy (Statistical Uncertainty: $\sigma_{Peak}$), from the the background counts at each energy (Statistical Uncertainty: $\sigma_{Bkg}$), and from the (n,$\alpha$) counts at 14.1 MeV used for normalization (Statistical uncertainties: $\sigma_{N_{n,\alpha}}$); the total statistical uncertainties combining all statistical uncertainties (Statistical Uncertainties: $\sigma_{stat. total}$); and systematic uncertainties, which include the charged particle detection efficiency correction uncertainty (Systematic Uncertainties: $\sigma_{Eff.}$), the relative flux distribution uncertainty (Systematic Uncertainties:  $\sigma_{Flux}$), the fit/integration uncertainties (Systematic Uncertainties: $\sigma_{Fit}$), the incident neutron energy reconstruction uncertainty (systematic uncertainties: $\sigma_{NKE}$), and the normalization uncertainty based on the weighted average uncertainty of previous data (Systematic Uncertainties: $\sigma_{Norm.}$). The relative uncertainty, which includes both statistical and systematic uncertainties, is included as a percentage, and the absolute uncertainty is included in millibarns.}
    \centering
    \begin{tabular}{|c|c|c|c|c|c|c|c|c|c|c|c|c|c|c|c|c|}
    \hline
$E_n$  & \multicolumn{2}{c|}{Counts} & \multicolumn{2}{c|}{Corrections (\%)} & Cross section & \multicolumn{4}{c|}{Statistical Uncertainties (\%)} &  \multicolumn{5}{c|}{Systematic Uncertainties (\%)} & Rel. Unc.(\%)& Abs. Unc.\\
\hline    
(MeV) & Peak & Total & Efficiency & Transmission & $\sigma$ (mb) & $\sigma_{Peak}$ &  $\sigma_{Bkg}$ &  $\sigma_{N_{n,\alpha}}$ & $\sigma_{stat. total}$ & $\sigma_{Eff.}$ & $\sigma_{Flux}$ & $\sigma_{Fit}$ & $\sigma_{NKE}$ & $\sigma_{Norm.}$ & $\sigma_{Rel.}$ & (mb)\\
\hline
   
17.78 & 176 & 1418 & 98.4 & 82.0 & 5.1 & 7.5 & 20.0 & 2.96 & 21.6 & 4 & 7.2 & 9.4 & 0.49 & 4.87 & 25.4 & 1.3 \\
18.28 & 109 & 891 & 97.3 & 81.4 & 3.2 & 9.6 & 25.6 & 2.96 & 27.5 & 4 & 7.4 & 13.3 & 0.49 & 4.87 & 32.1 & 1.0 \\
18.81 & 105 & 732 & 96.5 & 81.8 & 3.1 & 9.8 & 24.0 & 2.96 & 26.0 & 4 & 6.5 & 15.7 & 0.49 & 4.87 & 31.7 & 1.0 \\
19.35 & 136 & 671 & 95.0 & 81.0 & 4.1 & 8.6 & 17.0 & 2.96 & 19.2 & 4 & 5.4 & 10.4 & 0.49 & 4.87 & 23.4 & 1.0 \\
19.92 & 120 & 624 & 93.4 & 81.1 & 3.6 & 9.1 & 18.8 & 2.96 & 21.1 & 4 & 5.4 & 10.2 & 0.49 & 4.87 & 24.9 & 0.9 \\
20.52 & 132 & 536 & 91.5 & 81.7 & 4.0 & 8.7 & 15.2 & 2.96 & 17.8 & 4 & 5.7 & 9.8 & 0.49 & 4.87 & 22.0 & 0.9 \\
21.14 & 177 & 553 & 89.5 & 82.1 & 5.2 & 7.5 & 11.0 & 2.96 & 13.6 & 4 & 5.2 & 6.5 & 0.49 & 4.87 & 17.2 & 0.9 \\
21.80 & 166 & 622 & 87.7 & 82.5 & 5.0 & 7.8 & 12.9 & 2.96 & 15.3 & 4 & 6.1 & 7.2 & 0.49 & 4.87 & 19.0 & 0.9 \\
22.48 & 187 & 697 & 85.1 & 82.0 & 5.6 & 7.3 & 12.1 & 2.96 & 14.4 & 4 & 5.9 & 7.7 & 0.49 & 4.87 & 18.5 & 1.0 \\
23.20 & 97 & 329 & 82.2 & 82.2 & 2.9 & 10.1 & 15.6 & 2.96 & 18.8 & 4 & 6.9 & 7.4 & 0.49 & 4.87 & 22.3 & 0.6 \\

\hline
    \end{tabular}
\end{table}